\documentclass[conference]{IEEEtran}
\IEEEoverridecommandlockouts

\usepackage{cite}
\usepackage{amsmath,amssymb,amsfonts}
\usepackage{algorithmic}
\usepackage{graphicx}
\usepackage{textcomp}
\usepackage{xcolor}
\usepackage{float} 
\usepackage{booktabs} 
\usepackage{enumitem} 
\usepackage{placeins} 
\usepackage{array}    
\def\BibTeX{{\rm B\kern-.05em{\sc i\kern-.025em b}\kern-.08em
    T\kern-.1667em\lower.7ex\hbox{E}\kern-.125emX}}
\begin{document}

\title{\LARGE \bfseries Debatts: Zero-Shot Debating Text-to-Speech Synthesis}

\author{
    \begin{minipage}[t]{\textwidth}
        \centering
        Yiqiao Huang$^{1}$, Yuancheng Wang$^{1}$, Jiaqi Li$^{1}$, Haotian Guo$^{1}$, Haorui He$^{1}$, Shunsi Zhang$^{2}$, Zhizheng Wu$^{1}$\\
        $^{1}$The Chinese University of Hong Kong Shenzhen, China \\
        $^{2}$Guangzhou Quwan Network Technology \\
        \texttt{\
        yiqiaohuang@link.cuhk.edu.cn, wuzhizheng@cuhk.edu.cn} \\
    \end{minipage}
}

\maketitle

\begin{abstract}

In debating, rebuttal is one of the most critical stages, where a speaker addresses the arguments presented by the opposing side. During this process, the speaker synthesizes their own persuasive articulation given the context from the opposing side. This work proposes a novel zero-shot text-to-speech synthesis system for rebuttal, namely Debatts. Debatts takes two speech prompts, one from the opposing side (i.e. opponent) and one from the speaker. The prompt from the opponent is supposed to provide debating style prosody, and the prompt from the speaker provides identity information. In particular, we pretrain the Debatts system from in-the-wild dataset, and integrate an additional reference encoder to take debating prompt for style. In addition, we also create a debating dataset to develop Debatts. In this setting, Debatts can generate a debating-style speech in rebuttal for any voices. Experimental results confirm the effectiveness of the proposed system in comparison with the classic zero-shot TTS systems.
\end{abstract}



\section{Introduction}

\begin{figure}
    \centering
    \includegraphics[width=0.9\linewidth]{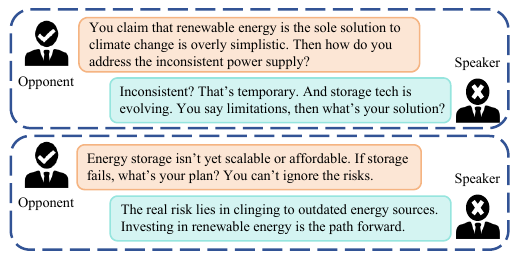}
    \caption{An illustration of rebuttals between two speakers. During rebuttal, the speaker's speaking style is impacted by the speech from the opponent. Words other than speech signal are used to represent speech debating for a better visualization. }
    \label{fig:rebuttal_example}
\end{figure}

Debating is a structured form of argument where participants engage on a specific topic, offering contrasting perspectives to persuade audiences or judges. A critical component of debating is the rebuttal, where a participant directly responds to the opponent's arguments or questions. Rebuttals demand persuasive articulation, as participants must both undermine their opponent’s claims and reinforce their own. Fig.~\ref{fig:rebuttal_example} illustrates of two rebuttal speeches responding two the opponent's questions. The effectiveness of a rebuttal often hinges on the debater's content and vocal delivery. Varied and expressive speech can enhance the clarity and impact of their points, ultimately influencing the outcome of the debate\cite{sen2021dbates}. Achieving naturalness and authenticity in vocal performance is vital to this process. \textit{This work explores the application of zero-shot Text-to-Speech (TTS) technology in the context of debating, with a specific focus on the rebuttal phase, where participants systematically challenge and counter their opponents' arguments}. 

\begin{figure*}
    \centering
    \includegraphics[width=0.9\linewidth]{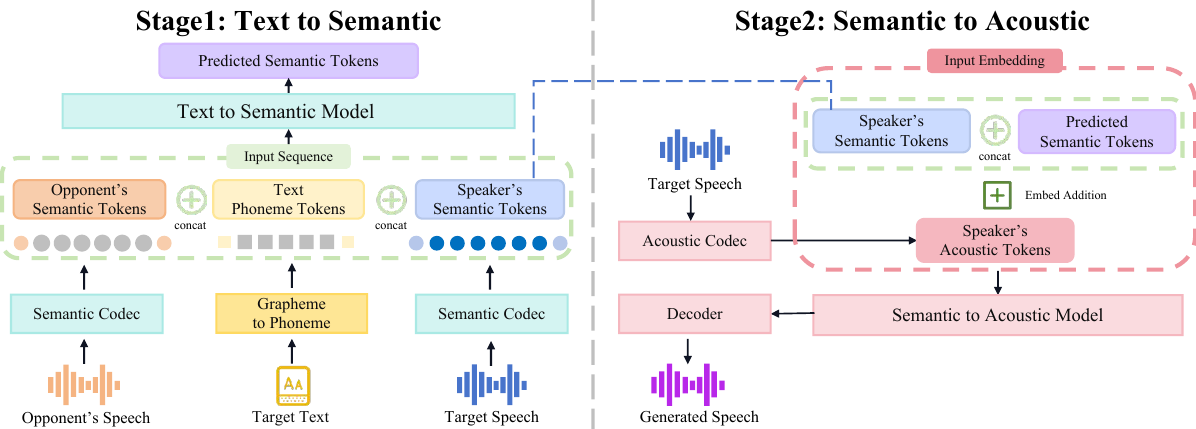}
    \caption{The proposed Debatts architecture. It consists of a text-to-semantic stage and a semantic-to-acoustic stage. In the first stage, the model predicts target semantic tokens using opponent and target speech semantic tokens along with text tokens. In the second stage, it generates speech with debating style based on the concatenated target speaker and predicted semantic tokens, along with the target speaker's acoustic tokens.}
    \label{fig:model architecture}
\end{figure*}



There are some studies on developing debating systems. To name a few, producing rebuttal in debating is defined as a natural language understand task in \cite{orbach2019rebuttal}. The study presents the problem of producing a rebuttal
response to a long argumentative text with a public dataset. In~\cite{slonim2021autonomous}, an autonomous debating system was developed with inspiration from the substantial progress in language model development. The debating system can engage in a competitive debate with humans. The existing studies mainly focus on the generation of debating text. 



Although text content of a rebuttal plays a critical role in debating, the vocal performance or prosody of the rebuttal speech is equally important (i.e. the rendering of rebuttal as a speech. To enhance the prosody of the generate speech, in \cite{mass2018word}, word emphasis prediction was proposed for expressive prosody generation. This work mainly focuses on the word-level emphasis prediction and only uses the target speaker's data for prosody and speech generation. 


As mentioned in~\cite{li2020emotions}, the speaker's rebuttal vocal performance is impacted by their own vocal characteristics and also the opponent's vocal performance (e.g. prosody, style). Inspired by the zero-shot text-to-speech synthesis~\cite{wang2023neural,borsos2023soundstorm, shen2023naturalspeech, ju2024naturalspeech, le2024voicebox} that takes speech prompt or  prompt of style instructions~\cite{skerry2018towards,min2021meta,huang2022generspeech,guo2021conversational,hu2022fctalker,xue2023m,xue2024improving}, we propose a zero-shot debating text-to-speech synthesis, namely Debatts. Debatts takes the opponent's speech as style or prosody prompt and the speaker's speech sample to speaker identity. In this way, Debatts is able to generate expressive rebuttal responses by considering the opponent's speech, and can generate rebuttal speech for a speaker who never debates. 

To support the development of Debatts system, we also create a multi-speaker debating dataset, namely Debatts-Data. Existing datasets for debating such as \cite{abercrombie2018sentiment,krishna2018analysis,ruiz2021vivesdebate}, focus mainly on text content. Some debate datasets such as~\cite{mirkin2017recorded,orbach2019rebuttal} are typically artificially produced and have limited reflection of the natural and spontaneous style of authentic debate scenarios. The Debatts-data dataset is sourced from formal Mandarin debates, offering invaluable resources for research in rebuttal generation. \textit{\textbf{To the best of our knowledge, this is the first Mandarin debating speech dataset and the first zero-shot TTS system that considers the opponent's speech as a prompt}}.

\section{Proposed Debatts System}



Debatts is built on top of the zero-shot TTS framework with inspirations from the previous work~\cite{borsos2023soundstorm,anastassiou2024seed,wang2024maskgct}. Debatts is a two-stage model as illustrated in Fig.~\ref{fig:model architecture}.
The first stage is to predict semantic tokens with opponent's speech, target text and target speech as inputs, while the second stage is to predict acoustic tokens which are then transformed into waveform.

\begin{figure*}
    \centering
    \includegraphics[width=0.9\linewidth]{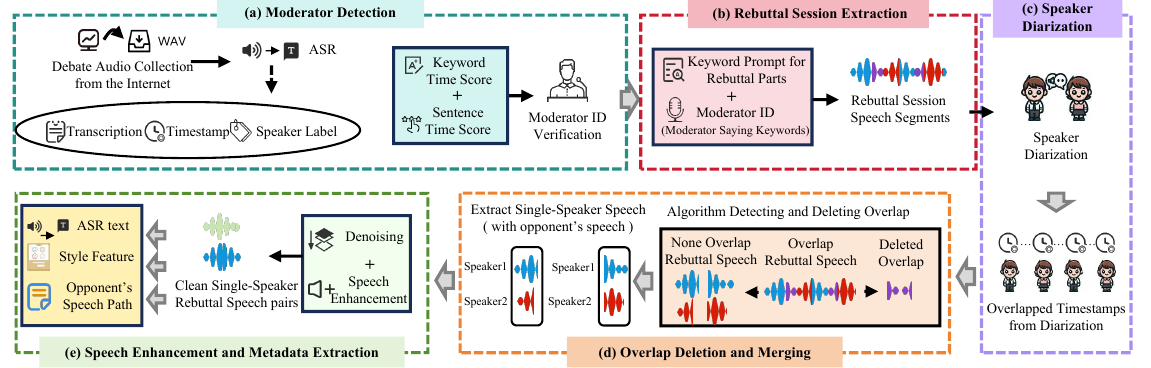}
    \caption{An illustration of the proposed five-step dataset pipeline to create the Debatts-data dataset. The pipeline consists of moderator detection, rebuttal session extraction, speaker diarization, overlap deletion and merging and speech enhancement and metadata extraction processes.}
    \label{fig:enter-label}
\end{figure*}

\subsection{Text-to-Semantic Prediction}
To generate semantic tokens, Debatts employs a LLaMA-style Transformer model~\cite{touvron2023llama} as the backbone. During pretraining, the speaker's prompt speech is processed through the semantic codec same as that in~\cite{wang2024maskgct}, resulting in a discrete semantic sequence $\boldsymbol{S_{spk}}$. Meanwhile, text semantic tokens $\boldsymbol{T}$ are generated from target text using a grapheme-to-phoneme tool. Based on the text sequence $\boldsymbol{T}$, the model autoregressively generates semantic tokens for the prompt.

During training, we incorporate the opponent’s speech as an additional input or condition for context learning. The opponent's speech is unified with the target reference speech at the discrete token level. Specifically, the opponent’s speech is passed through the same semantic codec to extract a discrete sequence $\boldsymbol{S_{op}}$. Then, $\boldsymbol{S_{op}}$, $\boldsymbol{T}$, and $\boldsymbol{S_{spk}}$ are concatenated as inputs, and model autoregressively predicts semantic tokens $\boldsymbol{\hat{S}_{spk}}$ with those inputs. The training process formulated as
\begin{equation}
\begin{aligned}
\text{$p(\hat{S}_{spk}|T,S_{op};\theta_{t2s}^{new})$} &= \text{$\prod_{t=1}^N p(\hat{S}_{spk,t}|S_{spk,<t},T,S_{op};\theta_{t2s}^{new})$}
\end{aligned}
\end{equation}
where $\boldsymbol{S_{spk}}$
denotes the target reference speech representation tokens sequence, $\boldsymbol{\hat{S}_{spk}}$ denotes the predicted tokens sequence, and $\boldsymbol{S_{op}}$ denotes the opponent's token sequence.

During training, the losses are calculated between $\boldsymbol{S_{spk}}$ 
 and $\boldsymbol{\hat{S}_{spk}}$, excluding $\boldsymbol{S_{op}}$ and $\boldsymbol{T}$ to ensure that the model only uses the opponent’s context as a condition. 
 The model then focuses on how prosody and emotional details of the opponent's speech shape the target speech style. As a result, it synthesizes debate style from the opponent's speech context and speaker timbre from the prompt, producing natural responses.


In the inference stage, we utilize the token sequences from opponent’s speech, target text, speaker’s prompt speech, and prompt text as conditions to generate the target speech.


\subsection{Semantic-to-Acoustic Generation}
To predict acoustic tokens from semantic tokens, we employ a model architecture similar to the SoundStorm model~\cite{borsos2023soundstorm}. The model takes the semantic tokens sequence $\boldsymbol{S}$ and the acoustic tokens sequence $\boldsymbol{A_{1:N}}$ from the prompt as input. The acoustic token prediction $\boldsymbol{A}$ consists of $\boldsymbol{N}$ layers. 
The acoustic tokens sequence of speaker prompt speech is extracted by a speech acoustic codec same as~\cite{wang2024maskgct}.
Since the frame numbers of $\boldsymbol{S}$ and $\boldsymbol{A_{1:N}}$ are equal, we simply add their embeddings and use them as the model input. During training, the model predict masked acoustic tokens based on semantic tokens and unmasked acoustic tokens.
In the inference stage, the model generates acoustic tokens layer by layer, starting from coarse acoustic information in the lower layers. It progressively generates more detailed information in the higher layers. The model concatenates the speaker semantic tokens and the predicted semantic tokens from text-to-semantic generation to form $\boldsymbol{S}$ as the input, while using the speaker acoustic tokens as $\boldsymbol{A}$.

\section{A Mandarin Chinese debating dataset}

We propose a Mandarin debating dataset called Debatts-data to support the development of Debatts. This section presents the statistics of the dataset and pipeline for dataset creation. Detailed statistics and comparison with existing debate datasets are presented in Table~\ref{table:dataset_statistics}. We note that different from the traditional TTS dataset, the rebuttal in debating has an opponent and a speaker.
Typically, the rebuttal session begins with a competition moderator introducing the process.

\begin{table}[htbp]
\centering
\footnotesize
\caption{Comparison of publicly Debate datasets.}
\label{table:dataset_statistics}
\renewcommand{\arraystretch}{1.2} 
\footnotesize 
\resizebox{\columnwidth}{!}{ 
\begin{tabular}{
  @{\hspace{0pt}}l@{\hspace{0.6pt}}  
  c@{\hspace{0.5pt}}                 
  >{\centering\arraybackslash}p{1cm}@{\hspace{0pt}} 
  >{\centering\arraybackslash}p{1cm}@{\hspace{0pt}} 
  c@{\hspace{1pt}}                 
  >{\centering\arraybackslash}p{1cm}@{\hspace{0pt}} 
  >{\centering\arraybackslash}p{1cm}@{\hspace{0pt}} 
  >{\centering\arraybackslash}p{1cm}@{\hspace{0pt}} 
}
\toprule
\textbf{Dataset} & \textbf{Lang} & \parbox{1cm}{\centering \textbf{Num} \\ \textbf{Spks}} & \parbox{1cm}{\centering \textbf{Dur} \\ \textbf{(hrs)}} & \parbox{1cm}{\centering \textbf{Text/} \\ \textbf{Speech}}  & \parbox{1cm}{\centering \textbf{SR} \\ \textbf{(kHz)}} & \parbox{1cm}{\centering \textbf{Wild/} \\ \textbf{Studio}} \\ 

\midrule
V.Debate\cite{ruiz2021vivesdebate}             & EN & -    & 24 (E.) & T & - & W \\
Record\cite{mirkin2017recorded}              & EN & 10  & 6 (E.) & T+S & 44.1 & S \\
Rebuttal\cite{orbach2019rebuttal}             & EN & 14  & 27 (E.)   & T+S & - & S \\
DBates\cite{sen2021dbates}             & EN & 140  & 70 (E.)   & T+S & 16 & W \\
\midrule
Debatts-data      & ZH & \textbf{2350} & \textbf{111.9}  & \textbf{T+S} & 16 & \textbf{W} \\
\bottomrule
\end{tabular}}

{
The number of speakers in Debatts is calculated by comparing speaker similarity within and across competitions. \textbf{"(E.)"} indicates total duration estimated from the paper, while \textbf{"-"} denotes unavailable information.}
\end{table}


We collected the debating recordings from Internet platforms (e.g. YouTube, Bilibili). The primary sources include official recordings from major events such as the ``International Mandarin Debate Invitational" and ``Mandarin Debate World Cup". Debating recordings between 2010 and 2024 are selected and cover more than 800 competitions of 1,200 hours.


Subsequently, we developed a pipeline to process the raw recordings. The pipeline consists of five steps as shown in Fig.~\ref{fig:enter-label}, including moderator detection, rebuttal session extraction, speaker diarization, overlap deletion and merging and speech enhancement and metadata extraction. In particular,
\begin{itemize}[left=0pt] 
\item \textbf{Moderator Detection: }
    First, we utilized Paraformer zh~\cite{anastassiou2024seed} to transcribe texts and extract global speaker labels and then identified the moderator's speaker ID as the anchor ID.
    \item \textbf{Rebuttal Session Extraction: }
    Then, we located the rebuttal session segments by extracting keywords from the moderators. This approach enables automated and precise extraction of specific segments from full competition recordings.
    \item \textbf{Speaker Diarization: }
    After that, we employed a speaker diarization toolkit \cite{Plaquet23,Bredin23} to identify speakers within the rebuttals and extract single-speaker speech pairs for training.
    \item \textbf{Overlap Detection and Merging: }
    Next, we eliminated overlaps, which are common in rebuttals, to ensure data quality. The detection of overlaps is achieved by analyzing the overlapping timestamps in the diarization data. After deleting the overlaps, the speech segments from the same speaker are concatenated.
    \item
    \textbf{Speech Enhancement and Metadata Extraction:}
    Finally, we performed speech enhancement and metadata extraction. Speech enhancement is mainly to reduce background noise and remove speech segments with severe noise. Metadata extraction is to extract transcriptions, speaker labels, timestamps and style vectors~\cite{ma2023emotion2vec}.
\end{itemize}
Through these processes, we produced the first Mandarin debating dataset for TTS. The dataset contains rich prosody and meta information. The pipeline can used for other languages and other speech domains with minor changes.

\begin{table*}[htbp]
\centering
\caption{Objective and subjective results comparing baseline, finetuned baseline and Debatts system for debating case using Debatts-data and for non-debating case using Emilia\cite{he2024emilia}. Note that the selected speeches only serve as speaker prompts, with the test utterance pairs unchanged}
\label{table:evaluation}
\begin{tabular}{l|c|c|c|c|c|c|c|c|c|c|c|c}
\hline\hline
 \textbf{} & \multicolumn{6}{c|}{\textbf{Debating Target Prompts}} & \multicolumn{6}{c}{\textbf{Non-Debating Target Prompts}} \\
\cline{2-13}
& \multicolumn{4}{c|}{Objective} & \multicolumn{2}{c|}{Subjective} & \multicolumn{4}{c|}{Objective} & \multicolumn{2}{c}{Subjective} \\
\hline
\parbox[c]{0.5cm}{\centering \textbf{Method}} & \rule{0pt}{0.4cm} \parbox[c]{0.6cm}{\centering WER↓} & \parbox[c]{0.6cm}{\centering SIM \\ {Spk↑}} & \parbox[c]{0.8cm}{\centering Style \\ Consist↑} & \parbox[c]{0.7cm}{\centering SIM \\ Style↑} & {\centering NMOS↑} & {\centering SMOS↑} & \parbox[c]{0.6cm}{\centering WER↓} & \parbox[c]{0.6cm}
{\centering SIM \\ Spk↑} & \parbox[c]{0.8cm}
{\centering Style \\Consist↑} & 
\parbox[c]{0.7cm}{\centering SIM \\ Style↑} & {\centering NMOS↑} & {\centering SMOS↑} \\
\hline
Groundtruth& \rule{0pt}{0.25cm}- & - & - & - & 4.83 $\pm$ 0.15 & - & - & - & - 
& - 
& 4.81 $\pm$ 0.13 & - \\
\hline
Baseline  & \rule{0pt}{0.27cm}15.62 & 87.13 & 42.64 & 75.54 & 3.76 $\pm$ 0.14 & 3.52 $\pm$ 0.10 & 15.04 
& 76.30
& 41.39 & 69.13 & 3.57 $\pm$ 0.10 & 2.93 $\pm$ 0.12 \\
\hline
Finetuned& \rule{0pt}{0.25cm}13.17 & 87.54 & 48.63 & 76.65 & 3.80 $\pm$ 0.14 & 3.59 $\pm$ 0.13 & 11.69 
& 82.31
& 43.71 & 72.23 & \textbf{3.61 $\pm$ 0.11}& 3.10 $\pm$ 0.10  \\
\hline
Debatts& \rule{0pt}{0.3cm}\textbf{11.69} & \textbf{87.68} & \textbf{51.69} & \textbf{78.77} & \textbf{3.81 $\pm$ 0.13} & \textbf{3.61 $\pm$ 0.10} & \textbf{11.02} 
& \textbf{83.27}
& \textbf{48.01} & \textbf{76.42} & 3.60 $\pm$ 0.09 & \textbf{3.18 $\pm$ 0.11} \\
\hline\hline
\end{tabular}
\end{table*}



\section{Experiments and Analysis}

\subsection{Experimental setup}



To evaluate the proposed model, we performed experiments using the Emilia dataset, which contains over 100,000 hours of multilingual speech\cite{he2024emilia}, and the Debatts-data dataset consisting of 101.09 hours of speech. We trained three models for comparison, namely,
\begin{itemize}[left=0pt]
    \item \textbf{Baseline:} We use a strong baseline with LLaMA-style auto-regressive (AR) to predict semantic token, and use the non-auto-regressive (NAR) SoundStorm model for acoustic prediction, similar to that in~\cite{wang2024maskgct}. The baseline model employs only the Emilia dataset for training, and performs zero-shot TTS on both dataset test sets. Note that the baseline system only take the target text and target speaker speech as input. 
    \item \textbf{Finetuned:} We finetuned the baseline model with the Debatts-data dataset. Note that the opponent's speech is not considered in the finetuned system.
    \item \textbf{Debatts:} The proposed model as illustrated in Fig.~\ref{fig:model architecture}. Both Emilia and Debatts-data datasets are used, and the opponent's speech is used as an additional prompt.
\end{itemize}

To assess the model's ability to transfer debating style and generate natural responses, we conduct ablation tests on unseen datasets with both debating and non-debating speaker prompts. In both cases, the test utterance pairs remain unchanged, with selected speech used solely as speaker prompt. In the debating case, we choose 329 prompts from Debatts-data test set, ensuring consistency by matching them with the same speaker and debate as the ground truth. For the non-debating case, we randomly select 329 non-debating prompts that without debating style from Emilia~\cite{he2024emilia}. 
Note that when using the non-debating prompt as the target speech, the opponent's speech is from the debating dataset, the target speech prompt is from the Emilia dataset (i.e. \textit{Emilia data only provides speaker identity information, expecting the debating style comes from the opponent's speech prompt}).

\subsection{Objective Evaluation}

We first evaluate the proposed system objectively and the results are presented in Table~\ref{table:evaluation}. We use word error rate (WER), speaker similarity (SIM Spk), style similarity (SIM style) and style consistency (Style Consist) for comparing prosody consistency between synthetic and ground-truth. WER is calculated by following~\cite{anastassiou2024seed}. The SIM Spk is the cosine similarity calculated using embedding from WavLM~\cite{chen2022wavlm}. The Style Consist is calculated between generated speeches and ground-truth to measure consistency with verified classification results from the Emotion2Vec~\cite{ma2023emotion2vec}. And the SIM Style is the cosine similarity between the style embedding from~\cite{ma2023emotion2vec}.

Without a surprise, the finetuned model outperforms the baseline across all objective metrics. This indicates that by utilizing the Debatts-data, the finetuned model handles long, expressive speech synthesis better in the debating scenario. 
Debatts outperforms both the baseline and finetuned models across the objective metrics. It suggests that by incorporating the opponent’s speech, Debatts can generate more natural and coherent debating speech. We want to highlight the dramantic improvement in style consistency and style similarity. Although the finetuned model can generate speech with debating style, Debatts can generate rebuttals with better prosody details. This can be reflected in style consistency via the style classification consistency and the sytle similarity. It confirms the effectiveness of Debatts, which takes the opponent's speech as an additional prompt.


We then evaluated whether Debatts can generate rebuttal speech for any voice. We use non-debating prompt as target speaker prompt to provide speaker identity information only. It is observed that Debatts can generate better rebuttal speech comparing with baseline and finetuned in terms of WER, style consistency and style similarity. It confirms that \textit{even with non-debating speaker prompts, the model consistently produces natural and expressive debating speech for the target speaker, showcasing its robustness}.


\subsection{Subjective Evaluation}
To further assess the naturalness and stylistic similarity of the generated speech, we conducted subjective evaluations using two five-scale mean opinion score as metrics: NMOS (naturalness MOS) and SMOS (style similarity MOS). For SMOS, we compare the style similarity to ground truth. We randomly selected 15 generated samples\footnote{Some samples are available: https://amphionspace.github.io/debatts} from each of the two evaluation cases and invited 20 native Mandarin speakers to participate. The results are based on a 90\% confidence interval. 

The results demonstrate that Debatts outperforms both the baseline and finetuned models. These findings confirm that incorporating the opponent’s contextual information, rich with complex prosodic and emotional cues, significantly enhances the generation of rebuttal speech. We then performed subjective evaluation on naturalness and similarity using Mean Opinion Scores (MOS). Note that the similarity evaluation mainly focuses on debating-style similarity, rather than speaker similarity. We asked 20 native Mandarin Chinese speakers to listen to the samples. Each listened to 15 samples for each evaluation set. The naturalness of the proposed system is on par with the baseline and finetuned systems. \textbf{The proposed system achieves significantly better similarity on debating style than the baseline, specially on the non-debating dataset.} Note that the opponent's speech is from the debating dataset and the target speech prompt is a non-debating prompt, which is expected to provide only speaker identify information.

\section{Conclusions}

This paper proposes a zero-shot debating text-to-speech synthesis system, Debatts. Debatts generates expressive rebuttal speech by integrating the opponent's speech as an additional prompt complementary to the target speaker prompt. Both objective and subjective results confirm the effectiveness of Debatts. With non-debating speech as the target speaker prompt, Debatts can generate debating speech for any voices. In the future, we will investigate style and prosody disentanglement for better performance and controllability.


\newpage

{\fontsize{12}{15}\selectfont
\bibliographystyle{IEEEtran}
\bibliography{references}
}



\end{document}